\documentclass[aps,prc,twocolumn,superscriptaddress]{revtex4-1}

\usepackage{bm}
\usepackage{hyperref}
\usepackage{amsmath,amssymb,amstext,amsthm,amsfonts,braket} 
\usepackage[pdftex]{graphicx} 
\usepackage [english]{babel}
\usepackage [autostyle, english = american]{csquotes}
\MakeOuterQuote{"}

\begin{document}


\title{\boldmath Measurement of beam asymmetry for $\pi^-\Delta^{++}$ photoproduction \\ on the proton at $E_\gamma$=8.5 GeV}

\affiliation{Arizona State University, Tempe, Arizona 85287, USA}
\affiliation{National and Kapodistrian University of Athens, 15771 Athens, Greece}
\affiliation{Carnegie Mellon University, Pittsburgh, Pennsylvania 15213, USA}
\affiliation{The Catholic University of America, Washington, D.C. 20064, USA}
\affiliation{University of Connecticut, Storrs, Connecticut 06269, USA}
\affiliation{Duke University, Durham, North Carolina 27708, USA}
\affiliation{Florida International University, Miami, Florida 33199, USA}
\affiliation{Florida State University, Tallahassee, Florida 32306, USA}
\affiliation{The George Washington University, Washington, D.C. 20052, USA}
\affiliation{University of Glasgow, Glasgow G12 8QQ, United Kingdom}
\affiliation{Goethe University Frankfurt, 60323 Frankfurt am Main, Germany}
\affiliation{GSI Helmholtzzentrum f\"ur Schwerionenforschung GmbH, D-64291 Darmstadt, Germany}
\affiliation{Institute of High Energy Physics, Beijing 100049, People's Republic of China}
\affiliation{Indiana University, Bloomington, Indiana 47405, USA}
\affiliation{Alikhanov Institute for Theoretical and Experimental Physics NRC Kurchatov Institute, Moscow 117218, Russia}
\affiliation{Thomas Jefferson National Accelerator Facility, Newport News, Virginia 23606, USA}
\affiliation{Forschungszentrum Juelich Nuclear Physics Institute, 52425 Juelich, Germany}
\affiliation{University of Massachusetts, Amherst, Massachusetts 01003, USA}
\affiliation{Massachusetts Institute of Technology, Cambridge, Massachusetts 02139, USA}
\affiliation{National Research Nuclear University Moscow Engineering Physics Institute, Moscow 115409, Russia}
\affiliation{Norfolk State University, Norfolk, Virginia 23504, USA}
\affiliation{North Carolina A\&T State University, Greensboro, North Carolina 27411, USA}
\affiliation{University of North Carolina at Wilmington, Wilmington, North Carolina 28403, USA}
\affiliation{Northwestern University, Evanston, Illinois 60208, USA}
\affiliation{Old Dominion University, Norfolk, Virginia 23529, USA}
\affiliation{University of Regina, Regina, Saskatchewan, Canada S4S 0A2}
\affiliation{Universidad T\'ecnica Federico Santa Mar\'ia, Casilla 110-V Valpara\'iso, Chile}
\affiliation{Tomsk State University, 634050 Tomsk, Russia}
\affiliation{Tomsk Polytechnic University, 634050 Tomsk, Russia}
\affiliation{A. I. Alikhanyan  National Science Laboratory (Yerevan Physics Institute), 0036 Yerevan, Armenia}
\affiliation{College of William and Mary, Williamsburg, Virginia 23185, USA}
\affiliation{Wuhan University, Wuhan, Hubei 430072, People's Republic of China}
\author{S.~Adhikari}
\affiliation{Old Dominion University, Norfolk, Virginia 23529, USA}
\author{C.~S.~Akondi}
\affiliation{Florida State University, Tallahassee, Florida 32306, USA}
\author{A.~Ali}
\affiliation{Goethe University Frankfurt, 60323 Frankfurt am Main, Germany}
\affiliation{GSI Helmholtzzentrum f\"ur Schwerionenforschung GmbH, D-64291 Darmstadt, Germany}
\author{M.~Amaryan}
\affiliation{Old Dominion University, Norfolk, Virginia 23529, USA}
\author{A.~Asaturyan}
\affiliation{A. I. Alikhanyan National Science Laboratory (Yerevan Physics Institute), 0036 Yerevan, Armenia}
\author{A.~Austregesilo}
\author{Z.~Baldwin}
\affiliation{Carnegie Mellon University, Pittsburgh, Pennsylvania 15213, USA}
\author{F.~Barbosa}
\affiliation{Thomas Jefferson National Accelerator Facility, Newport News, Virginia 23606, USA}
\author{J.~Barlow}
\author{E.~Barriga}
\affiliation{Florida State University, Tallahassee, Florida 32306, USA}
% ...
\collaboration{The \textsc{GlueX} Collaboration}


\begin{abstract}
We report a measurement of the $\pi^-$ photoproduction beam asymmetry for the reaction $\vec{\gamma} p \rightarrow \pi^- \Delta^{++}$ using data from the \textsc{GlueX} experiment in the photon beam energy range 8.2--8.8~GeV. The asymmetry $\Sigma$ is measured as a function of four-momentum transfer $t$ to the $\Delta^{++}$ and compared to phenomenological models. We find that $\Sigma$ varies as a function of $t$:  negative at smaller values and positive at higher values of $|t|$. The reaction can be described theoretically by $t$-channel particle exchange requiring pseudoscalar, vector, and tensor intermediaries. In particular, this reaction requires charge exchange, allowing us to probe pion exchange and the significance of higher-order corrections to one-pion exchange at low momentum transfer. Constraining production mechanisms of conventional mesons may aid in the search for and study of unconventional mesons. This is the first measurement of the process at this energy.
\end{abstract}

\maketitle

\section{INTRODUCTION}
Determining the types of mesons that emerge from quantum chromodynamics (QCD) is a critical experimental input to our understanding of how QCD generates the properties of hadrons~\cite{Shepherd:2016dni}.  The \textsc{GlueX} experiment at Jefferson Lab provides a unique opportunity to search for non-$q\bar{q}$ mesons and, by using a linearly polarized photon beam, study their production dynamics in addition to their decay properties.  The \textsc{GlueX} photon beam energy of 8-9~GeV is in a regime where photoproduction of hadrons can be described by $t$-channel exchange processes~\cite{REGGE_PHENOMONOLOGY}, and the properties of exchanged Reggeons can be constrained by experimental data.  In particular, the linear polarization of the beam allows one to distinguish between exchange of particles with natural ($P(-1)^J=1$) and unnatural ($P(-1)^J=-1$) parity~ \cite{Stichel_theorem,Stichel_theorem_extended}.  Ultimately, this gives insight into the coupling of the produced meson and the photon to particular sets of Reggeons.  This knowledge of production mechanisms for known mesons can be leveraged in the future search for exotic hybrid mesons using \textsc{GlueX} data.

Measurements that constrain production mechanisms at photon beam energies relevant for the \textsc{GlueX} experiment are sparse. Recent measurements on the photoproduction of pseudoscalar mesons~\cite{gluex_pi0_eta,gluex_eta_etaprime,gluex_kplus_sigma} have begun to provide insight into into production mechanisms. In this paper, we seek to extend this understanding by measuring the beam asymmetry $\Sigma$ for the charge-exchange reaction $\vec{\gamma} p \rightarrow \pi^- \Delta^{++}$, where $\Sigma=1$ ($\Sigma=-1$) is indicative of pure natural (unnatural) parity exchange. We find that the asymmetry varies significantly over Mandelstam $t$, demonstrating the need for unnatural pion exchange as well as natural exchanges such as $\rho$ and $a_2$. This reaction has been of theoretical interest for several decades~\cite{goldstein_piDelta,clark_piDelta,campbell_piDelta}; however, most prior measurements have been made at lower energies \cite{SLAC_asymm_1970,SLAC_asymm_1972,LEPS_asymm,DESY_SDMEs,ABBHHM_SDMEs}. At energies of $E_\gamma=1-4$~GeV, both $t$-channel and $s$-channel processes contribute to single pseudoscalar photoproduction, and often the experimental focus is on $s$-channel baryon resonances.  Our measurements at higher energy will constrain the $t$-channel background for these investigations.

We report the first measurement of beam asymmetry $\Sigma$ for $\pi^-$ photoproduction at 8.5~GeV.  The analysis utilizes 20~pb$^{-1}$ of data collected by the \textsc{GlueX} experiment in 2017 at the Hall~D facility.  We compare our results to theoretical predictions at $E_\gamma$=8.5~GeV provided by the JPAC Collaboration~\cite{jannes_piDelta} and B.-G.~Yu and K.-J.~Kong~\cite{yu_piDelta}. These models are informed by cross section and asymmetry results for this reaction measured at $E_\gamma$=16~GeV with data from SLAC~\cite{SLAC16GeV_asymm}, the only previous measurement in this energy regime. 

\section{EXPERIMENTAL APPARATUS}

The \textsc{GlueX} experiment utilizes the 12~GeV Continuous Electron Beam Accelerator Facility (CEBAF) to produce a beam of linearly polarized photons via coherent bremsstrahlung radiation on a thin (50~$\mu$m) diamond wafer \cite{gluex_nim}. Measuring the momentum of the electron after radiation using a hodoscope allows the energy of the radiated photon to be determined with a resolution of 10~MeV in the beam energy range of interest.  By orienting the radiator, one may tune the coherent bremsstrahlung peak energy and direction of linear polarization. Four data sets of approximately equal statistics were collected with the coherent bremsstrahlung enhancement in the 8.2--8.8~GeV region and polarization oriented in four directions relative to the laboratory floor plane: -45$^\circ$, 0$^\circ$, 45$^\circ$, and 90$^\circ$.  We group these independent data sets in pairs of orthogonal orientations and refer to them as `0/90' or `-45/45', each of which is used to make a measurement of the observable of interest.  Within each set we label the 0 and -45 as $\parallel$ and the 90 and 45 as $\perp$.  

\begin{figure}
	\includegraphics[width=0.95\linewidth]{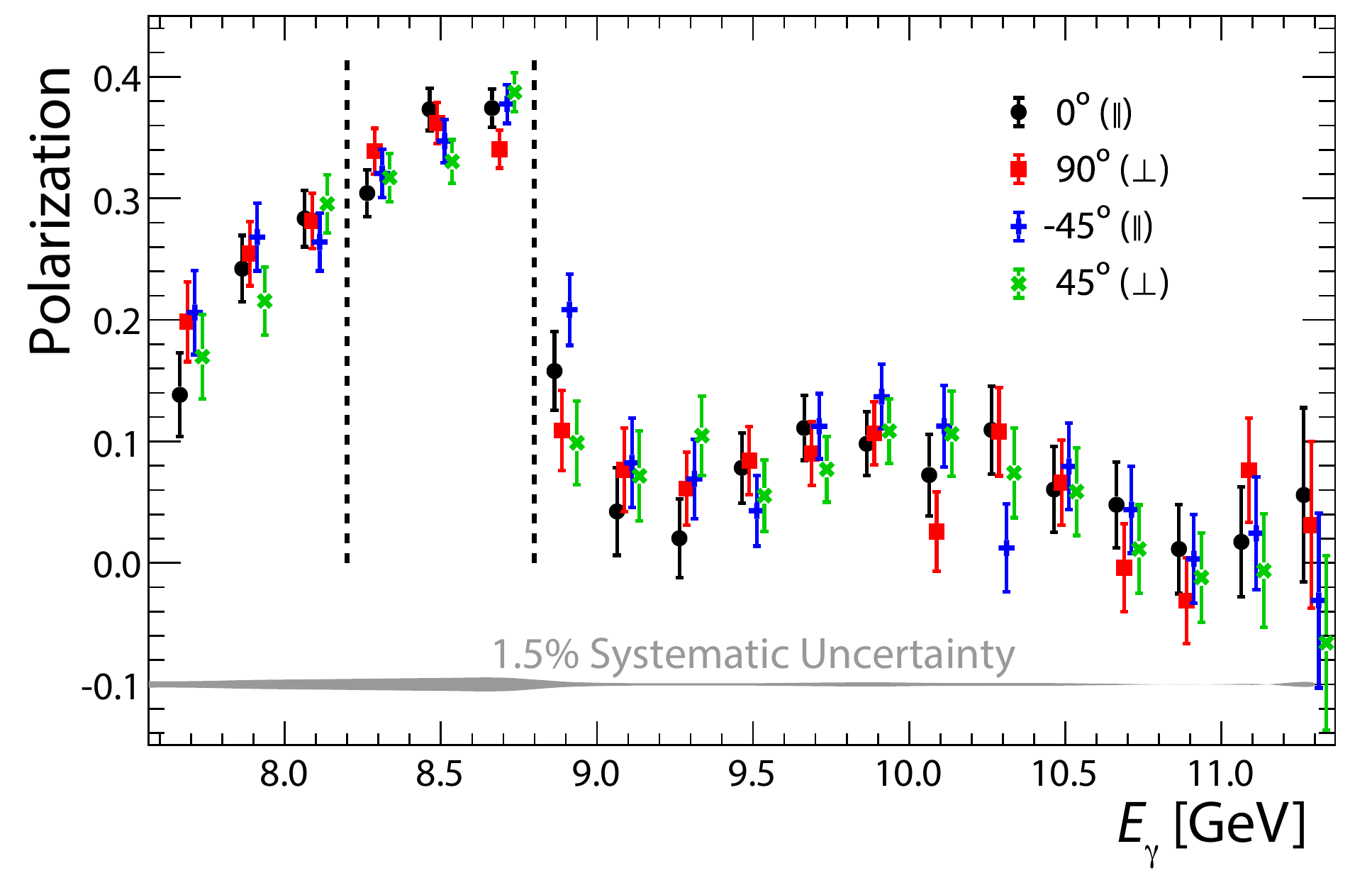} 
	\caption{
		The degree of linear polarization for four different orientations of diamond radiator as a function of beam photon energy, as measured by the TPOL. Events between the dashed lines (8.2~GeV$<\!E_\gamma\!<$8.8~GeV) are analyzed. (Data points are slightly offset for clarity.)
	}
	\label{fig:polarization}
\end{figure}

Beam photons travel 75~m from the radiator and pass through a 5~mm diameter collimator to enhance the polarization, as coherent bremsstrahlung photons are preferentially produced at small angles with respect to the beam axis. A downstream 75~$\mu$m beryllium converter allows for photon beam flux and polarization measurements. Flux is measured from $e^+e^-$ pair production measured in a pair spectrometer (PS)~\cite{NIM_PS}. Polarization is measured via detection of the recoil atomic electron of the triplet production process in the triplet polarimeter (TPOL)~\cite{NIM_TPOL}.  The azimuthal angle of this electron is sensitive to the photon polarization plane.  The photon polarization is measured independently for each polarization direction as a function of $E_\gamma$, with polarization values up to 40\%, as shown in Fig.~\ref{fig:polarization}. The statistical uncertainty in polarization is determined by the number of triplet production events detected. The systematic uncertainty of the instrument is 1.5\%.

The \textsc{GlueX} spectrometer is an azimuthally symmetric detector located in Hall~D of Jefferson Lab. The central elements of the detector are housed in a 2~T superconducting solenoid. Incident beam photons interact in a 30~cm long target filled with liquid hydrogen. The target is surrounded by the Start Counter (ST)~\cite{NIM_SC}, a scintillating detector which provides determination of the primary interaction time and allows for matching to radiating electrons in the upstream tagger.

Charged particles exiting the target are measured by two drift chamber systems: the Central Drift Chamber (CDC) ~\cite{NIM_CDC_old,NIM_CDC_new} and the Forward Drift Chamber (FDC)~\cite{NIM_FDC,FDC_IET}. The CDC consists of 28 layers of straw tubes surrounding the target region arranged in stereo and axial layers, providing track reconstruction to beyond 120$^\circ$ and allowing for proton-pion separation below about 1~GeV/$c$ based on energy loss ($dE/dx$). The FDC, located immediately downstream, consists of four planar packages. Each FDC package contains anode wire and cathode strip readouts. These two tracking systems allow for charged track reconstruction with uniform azimuthal coverage, polar angle coverage from 1$^\circ$ to beyond 120$^\circ$, and a momentum resolution of about 1-7\% depending on momentum and direction. In the forward direction, a Time-of-Flight (TOF) scintillator wall \cite{NIM_TOF_first,NIM_TOF_second} provides additional charged particle timing information.

Photon detection with the \textsc{GlueX} spectrometer is performed with two distinct electromagnetic calorimeters: the Barrel Calorimeter (BCAL)~\cite{NIM_BCAL} and the Forward Calorimeter (FCAL)~\cite{NIM_FCAL}. The BCAL surrounds the two drift chambers and is composed of 48 azimuthal lead-scintillating fiber matrix segments. The BCAL provides polar angle coverage from $11^\circ$ to $120^\circ$. The FCAL is located approximately 6~m downstream from the target and consists of 2,800 lead-glass blocks in a circular arrangement, providing azimuthally symmetric coverage for polar angles $1^\circ$ to $11^\circ$. Detector readout is triggered based upon energy deposition in the two calorimeters. 

\section{EVENT SELECTION}

We detect the $\Delta^{++}$ baryon via its dominant decay \mbox{$\Delta^{++} \rightarrow \pi^+p$}, hence we reconstruct the final state \mbox{$\vec{\gamma} p \rightarrow \pi^+ \pi^- p$}. A beam energy satisfying $8.2$~GeV$<E_\gamma<8.8$~GeV is required to select a sample of events with a high degree of linear polarization. Exactly two positive tracks and one negative track are required during reconstruction. We require that tracks originate from the target volume and produce hits in the TOF or BCAL. Both detectors provide timing information used for time-of-flight measurements, which are required to be consistent with either a proton or pion hypothesis, as appropriate. The vast majority of protons in this topology are produced at polar angles greater than 20$^\circ$ and with momentum lower than 1~GeV/$c$. In this case, energy loss $dE/dx$ measured in the CDC is effective at further distinguishing proton and $\pi^+$ candidates.

Each reconstructed event is also required to be matched to a suitable reconstructed radiating electron that is a candidate for the electron that radiated the beam photon.  The momentum of this electron determines the photon energy. The CEBAF accelerator delivers the electron beam in bunches with a 4~ns period. Hit information from the ST determines the beam bunch, and a precise value of arrival time of the bunch at the target center ($t_{\text{bunch}}$) is provided by the accelerator radio-frequency clock. We require electron candidates have a time $t_e$ such that $|t_e-t_{\text{bunch}}|<2$~ns. Due to the hit multiplicity in the tagger, more than one electron is typically detected per event, though only one of these electrons corresponds to the beam photon that interacted downstream. To remove electrons incorrectly (``accidentally'') associated with the triggered downstream event, we also select a statistically independent sample of events that satisfy $2$~ns$~<|t_e-t_{\text{bunch}}|<18$~ns. This selects eight additional beam bunches which, when scaled appropriately, can be used to remove the contribution of these accidentals to the analysis.

We impose several constraints to ensure the purity of the exclusive reaction of interest. First, the measured missing mass squared is required to satisfy $|p_{\text{i}}-p_{\text{f}}|^2<0.1$~GeV$^2$ to suppress the contribution from events with undetected massive particles, where $p_{\text{i}}$ and $p_{\text{f}}$ are the sum of all initial and final four-momenta respectively. Then, a kinematic fit is performed, enforcing conservation of energy and momentum and a common vertex, assuming the exclusive topology $\vec{\gamma} p \rightarrow \pi^+ \pi^- p$. We require that the kinematic fit $\chi^2$ satisfies $\chi^2/\text{NDF}\!<\!8.7$ to ensure that events are well-reconstructed and match the desired topology~\cite{JZ_thesis}. 

\begin{figure}
	\includegraphics[width=\linewidth]{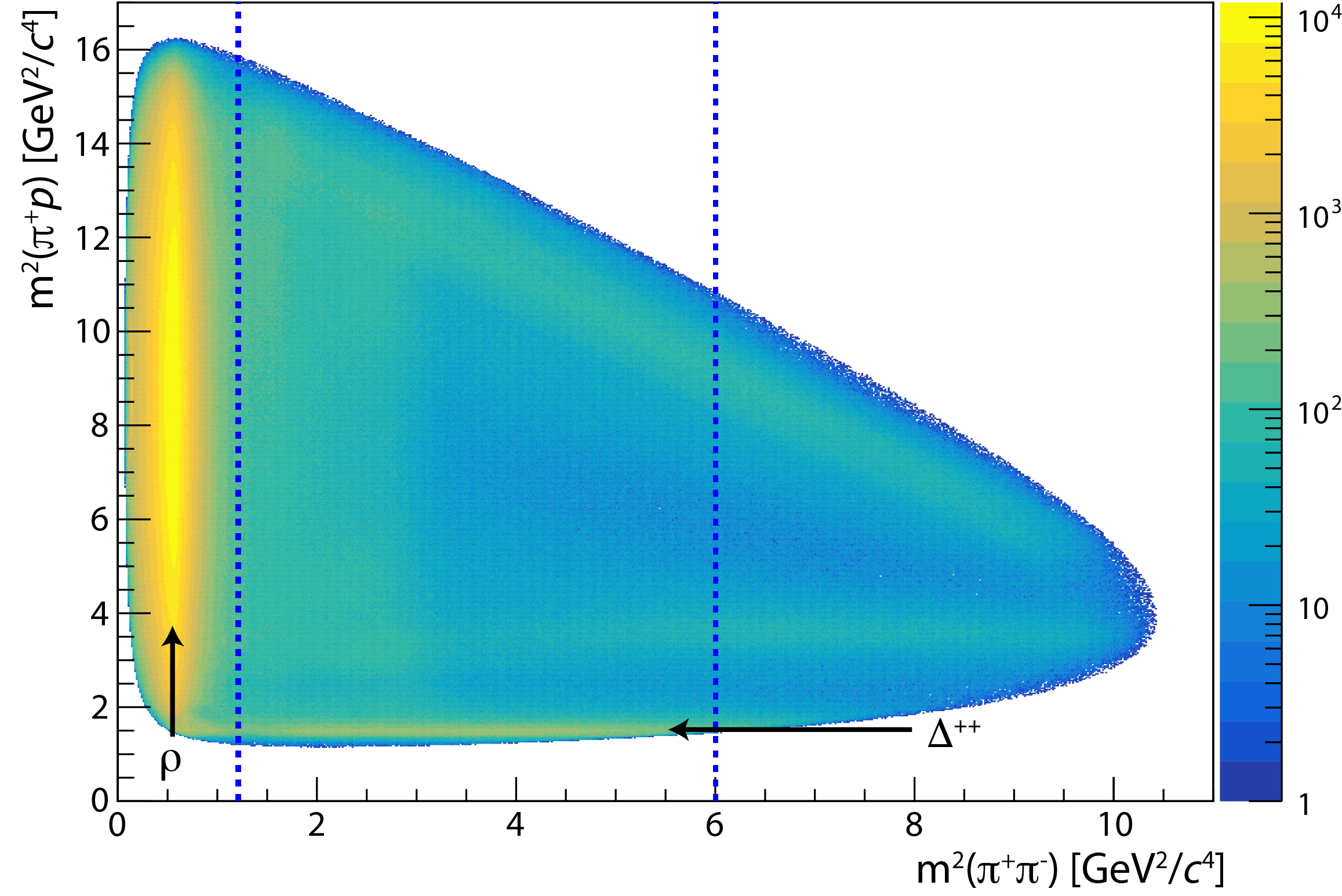}
	\caption{
		Dalitz plot of products of the reaction $\vec{\gamma}p\rightarrow\pi^+\pi^-p$. Candidates between dashed lines are selected. Data shown are not efficiency corrected.
	}
	\label{fig:dalitz}
\end{figure}

A number of intermediate states contribute to the reaction $\vec{\gamma} p \rightarrow \pi^+ \pi^- p$ in addition to the desired $\pi^-\Delta^{++}$ channel. In particular, the topology is dominated by production of the $\rho^0$ meson. We require $1.10~$GeV$/c^2<m_{\pi^+\pi^-}<2.45$~GeV$/c^2$ to reduce backgrounds, particularly from $\rho$ and $\Delta^*$ production.  This selection removes most of the $\rho^0$ background, as shown in Fig.~\ref{fig:dalitz}.  The production of $\Delta^0$ is also visible as a diagonal band in Fig.~\ref{fig:dalitz}; however, this process is well-separated from the signal region in $m_{p\pi^+}$.

\section{ANALYSIS}

The differential cross section for pseudoscalar production by a polarized photon beam is related to the total cross section $\sigma_0$ by
\begin{equation}
\label{eq:pol_xsec}
\frac{d\sigma}{d\phi}  =  \frac{\sigma_0}{2\pi} \Big{(} 1 - P_\gamma \ \Sigma\ \cos\left[2(\phi - \phi_{\text{lin}})\right]\Big{)},
\end{equation}
where $\phi$ is the azimuthal angle of the production plane in the lab, $\phi_{\text{lin}}$ is the azimuthal angle of beam polarization in the lab, $P_\gamma$ is the degree of linear polarization of the beam, and $\Sigma$ is the observable to be measured~\cite{photoprod_observables}. By using data collected with linear polarization in orthogonal directions, the term $\Sigma$ can be isolated without explicitly determining the total cross section or any $\phi$-dependent detector acceptance.  

\begin{figure}
	\includegraphics[width=\linewidth]{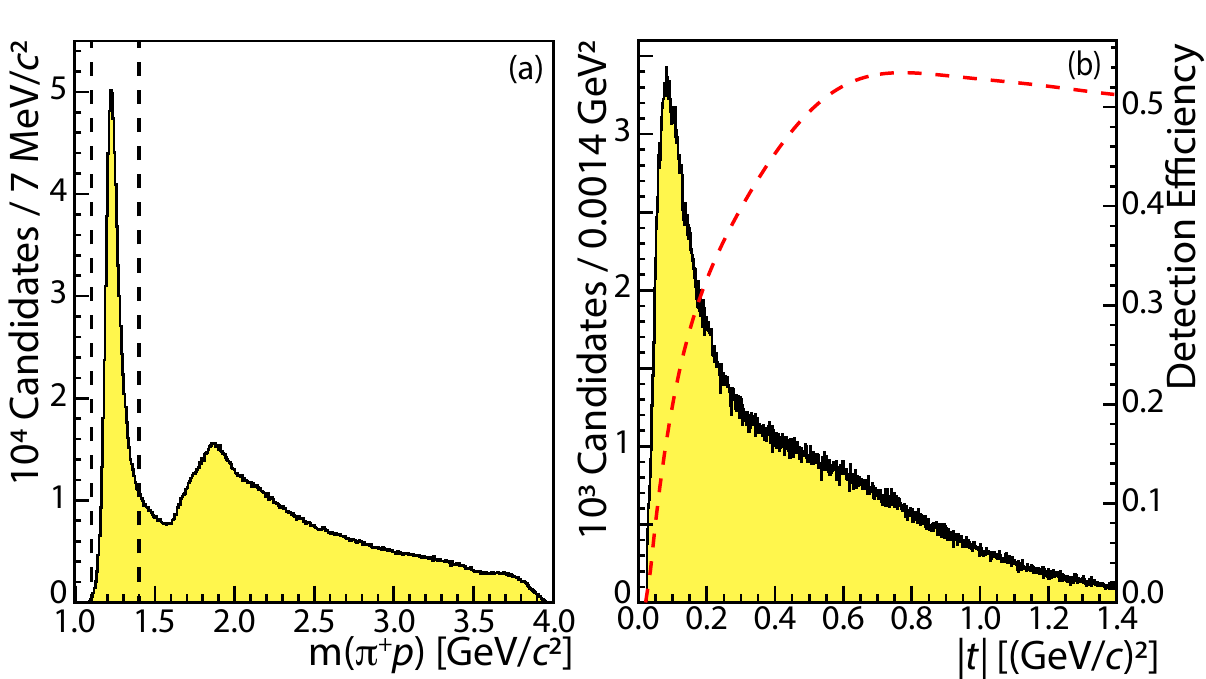} 
	\caption{
		(a) The $\pi^+p$ invariant mass distribution of events satisfying all selection criteria.  In addition to the  $\Delta^{++}$, excited states around 1.9~GeV/$c^2$ are visible.  (b) The distribution of $|t|$ for candidates between the dashed lines in panel (a) and the detection efficiency as a function of $|t|$. 
		Data plotted are not efficiency corrected. \label{fig:invmass_t_dist}
		}
\end{figure}

As shown in Fig.~\ref{fig:invmass_t_dist}(a), selecting a region of $m(\pi^+p)$ invariant mass does not ensure a pure sample of $\Delta^{++}$ events. Previous analyses typically first select a pure sample of events, and then produce a distribution in $\Delta\phi$. (Here $\Delta\phi \equiv \phi - \phi_\mathrm{lin}$ in Eq.~\ref{eq:pol_xsec}.) The amplitude of the $\cos(2\Delta\phi)$ component is then extracted to obtain $\Sigma$. In what follows, we perform the steps in reverse order: we project the $\cos(2\Delta\phi)$ component of all data and then isolate the $\Delta^{++}$ contribution by using the known lineshape of the $\Delta^{++}$.  The technique follows from that used to determine coefficients of a Fourier expansion. One can weight individual events by $\cos(n\Delta\phi)$ (where $n$ is an even integer) to create weighted histograms in $m(\pi^+p)$, thereby integrating over $\Delta\phi$.  The bin-by-bin contents of such histograms are then proportional to the strength of the $\cos(n\Delta\phi)$ component.  One can then fit these histograms, referred to later as $H_n$, to measure the $\Delta^{++}$ contribution to each, referred to as $Y_n$, with the $Y_2$ component being most sensitive to $\Sigma$. Practically, one must use orthogonal orientations of the beam polarization to cancel detector acceptance in the formulation of $\Sigma$. The full prescription for implementing this technique is documented in Refs.~\cite{dugger_moment_asymm,JZ_thesis}.

Following this prescription, we define a set of weighted invariant mass $m(\pi^+p)$ histograms for each separate orientation of polarization $H_n^{\perp/\parallel}$, each with accidental beam photon candidates subtracted as described above. Data are given an event-by-event weighting of $\cos(n\Delta\phi)$, using $\phi_{\text{lin}}$ as appropriate for each orientation of the beam polarization. The shape of the $\Delta^{++}$ in each $t$ region can be described by a relativistic Breit-Wigner function multiplied by a phase space factor~\footnote{M.~Mikhasenko, private communication.}:
\begin{equation} \label{eq:sig_function}
S(m) = \frac{|\mathbf{p}|}{m} \left| \frac{A}{m_0^2-m^2-im\Gamma(m) } \right|^2,
\end{equation}
where $A$ is a parameter determined by a maximum likelihood fit, and 
\begin{equation}
\Gamma(m)=\Gamma_0 \left(\dfrac{m_0}{m} \right) \left(\dfrac{|\mathbf{p}|}{|\mathbf{p}_0|} \right)^3 \left(\dfrac{1+|\mathbf{p}|^2a^2}{1+|\mathbf{p}_0|^2a^2} \right).
\end{equation}
Here, $m$ and $\mathbf{p}$ refer to the invariant mass of the $\pi^+ p$ system and the three-momentum of the proton (or pion) in the $\pi^+ p$ rest frame.  The  values of $m_0$ and $\Gamma_0$ are $\Delta^{++}$ resonance parameters obtained from Ref.~\cite{pdg}, and $|\mathbf{p}_0|$ is $|\mathbf{p}|$ computed at $m=m_0$.  The interaction radius $a$ is taken from Ref.~\cite{PEDRONI_HADROPRODUCTION}. Thus, the signal component of the fit contains a single free parameter $A$ in the equation above.  We use a fourth order Bernstein polynomial set to describe the smoothly varying background in the $m(\pi^+ p)$ spectra.  By integrating the signal fit function, we extract the moment-weighted yield of $\Delta^{++}$ candidates $Y_n$ corresponding to a particular histogram $H_n$.

Following Ref.~\cite{dugger_moment_asymm}, $\Sigma$ can then be expressed as 
\begin{equation}\label{eq:sig_mom}
\Sigma = \frac{Y_2^\perp+F_R Y_2^\parallel}{ \frac{P_\parallel}{2}(Y_0^\perp+Y_4^\perp)+\frac{F_R  P_\perp}{2}(Y_0^\parallel+Y_4^\parallel) },
\end{equation}
where $P_\perp$($P_\parallel)$ is photon polarization in $\perp$($\parallel$) datasets, and $F_R=N_\perp/N_\parallel$ is the ratio of measured photon flux for these data sets~\footnote{M. Dugger \textit{et al.} uses absolute angle $\phi$ rather than angle relative to polarization direction $\Delta\phi$, leading to a sign difference in term $F_R Y_2^\parallel$ of Eq.~\ref{eq:sig_mom}}.  While the \textsc{GlueX} detector was designed to be uniform in $\phi$, this need not be assumed:  any non-uniform azimuthal acceptance effects are removed by taking the difference of two orthogonal polarization directions and by including the terms $Y_4^\perp$ and $Y_4^\parallel$.

\begin{figure}
	\includegraphics[width=\linewidth]{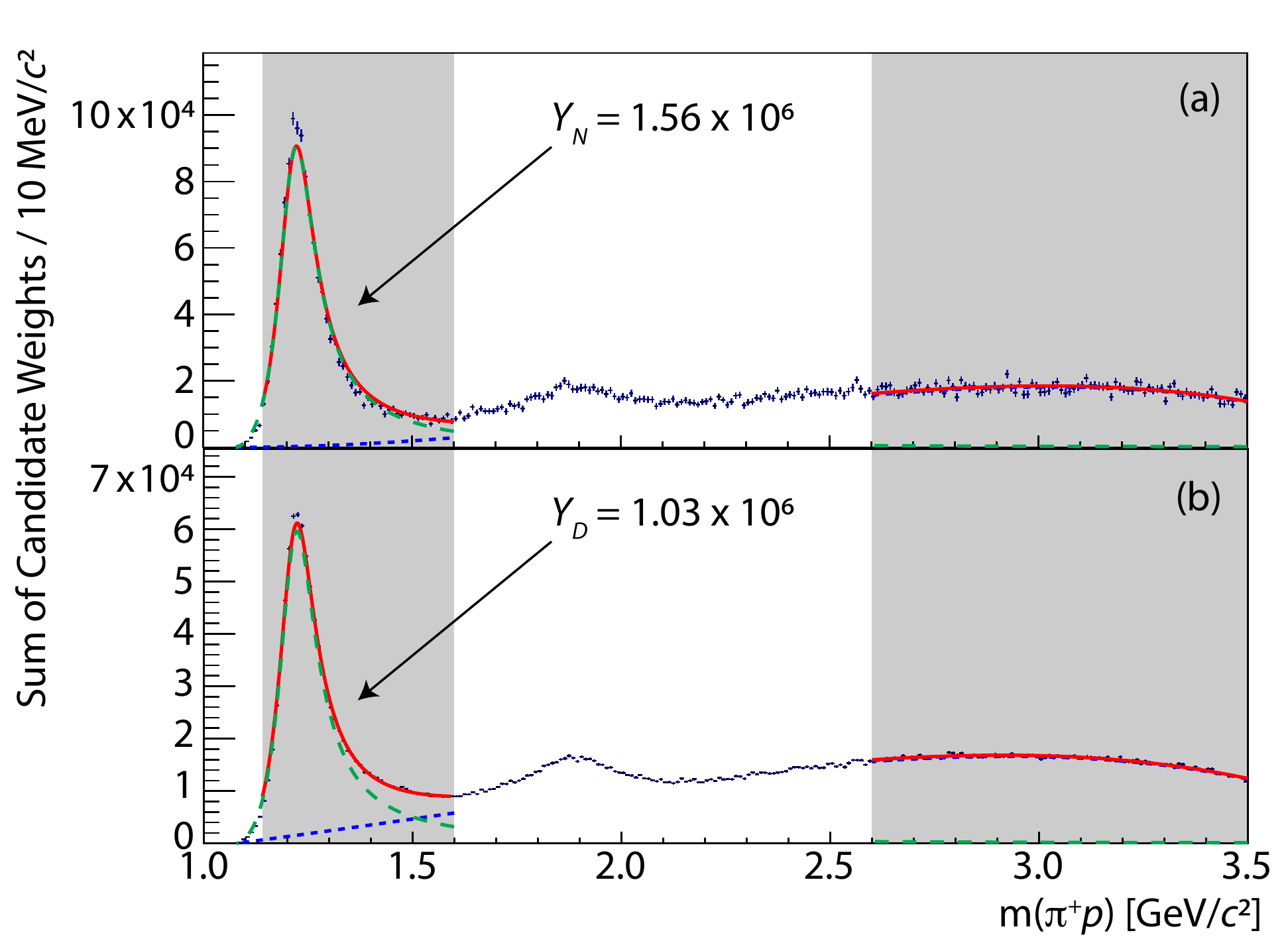} 
	\caption{
		(color online) Fit to (a) numerator $N$ and (b) denominator $D$ defined in Eqs.~\ref{eq:num} and~\ref{eq:den}, in the extended range $0.4$~(GeV/$c)^2<|t|<1.4$~(GeV/$c$)$^2$.  The $\Delta^{++}$ component is shown in green (dashed), polynomial background in blue (dotted), and total fit in red. Data are fit in the shaded regions only, the integral of the green (dashed) curve in the lower shaded region is used to determine the yields $Y_N$ and $Y_D$.
	}
	\label{fig:numpr_example}
\end{figure}

In practice, rather than fit each individual histogram $H_n^\perp$ and $H_n^\parallel$ to extract the $Y_n$, we note that the numerator and denominator in Eq.~\ref{eq:sig_mom} are linear combinations of terms $Y_n$, and hence we can construct two histograms $D$ and $N$, where the contents of the $i^{th}$ mass bin for each histogram (denoted $D_i$ and $N_i$) are given by the linear combinations
\begin{subequations}
	\begin{align}
	D_i & = \frac{P_\parallel}{2}(H_{0,i}^\perp+H_{4,i}^\perp)+\frac{F_R  P_\perp}{2}(H_{0,i}^\parallel+H_{4,i}^\parallel), \label{eq:den} \\
    N_i & = H_{2,i}^\perp+F_R H_{2,i}^\parallel + D_i. \label{eq:num}
	\end{align}
\end{subequations}
Let the weighted yield of $\Delta^{++}$ events in histograms $N$ and $D$ be denoted as $Y_N$ and $Y_D$ respectively. In terms of these two quantities, the asymmetry is then given by
\begin{equation}\label{eq:sigma_ND}
\Sigma = \dfrac{Y_N}{Y_D} - 1.
\end{equation}
In this formulation, $Y_N$ and $Y_D$ must be positive in order to be physical. This is advantageous, as likelihood fitting techniques can then be employed. 
We use this method to fit the $m(\pi^+ p)$ spectrum in the mass ranges $1.14$~GeV/$c^2<m(\pi^+ p)<1.60$~GeV/$c^2$ and $2.60$~GeV$/c^2<m(\pi^+ p)<3.50$~GeV/$c^2$, where the lower mass region contains the majority of the $\Delta^{++}$ signal and the higher mass region is used to further constrain backgrounds while avoiding $\Delta^*$ contributions. Figure~\ref{fig:numpr_example} shows a fit to $N$ and $D$ histograms obtained over a large $t$ range to demonstrate the ability of the lineshape to describe the data at high statistical precision.  Data are segmented into 16 regions of $t$, and in each region the 0/90 and -45/45 data sets provide two independent measurements of $\Sigma$.

The triply-differential cross section that describes the production of the $\Delta^{++}$ in each bin of $|t|$ can be written in terms of spin density matrix elements $\rho^\alpha_{\lambda \lambda^\prime}$ (SDMEs). Here $\lambda$ and $\lambda^\prime$ are helicities of the initial and final states.  The index $\alpha$ runs over the four spacetime dimensions and is contracted with the beam polarization four-vector in the expression for the triply-differential cross section.  When the two angles related to the polarization of the $\Delta^{++}$ are integrated over, one obtains the expression in Eq.~\ref{eq:pol_xsec} with $\Sigma = 2 \left[ \rho^1_{33} + \rho^1_{11} \right]$, where $\rho^\alpha_{\lambda \lambda^\prime}$ are SDMEs as defined in Ref.~\cite{Yu:2017kng}.  Experimentally, the non-uniform efficiency of detecting the $\Delta^{++}$ decay results in a weighted integration over the decay phase space.  This leads to a non-equal weighting of $\rho^1_{33}$ and $\rho^1_{11}$ and the introduction of other SDMEs that may cause the measured value of $\Sigma$ to deviate from the above expression. To correct for this bias, we use a \textsc{Geant4}~\cite{Allison:2016lfl} Monte Carlo (MC) simulation to calculate the efficiency $\epsilon$ as a function of the two decay angles in the $\Delta^{++}$ rest frame for each bin of $|t|$.  We then introduce an additional event-by-event weight of 1/$\epsilon$ down to a cutoff value of $\epsilon = 0.1\%$.  We exclude events in regions of phase space with efficiency lower than this.  Averaged over all bins of $|t|$, the effect of this weighting modifies $\Sigma$ by a magnitude of about 40\% of its total uncertainty. After this procedure, we find any residual bias to be negligible.  Separately, we use MC simulation to evaluate $m(\pi^+ p)$ and $t$ dependent modifications to the $\Delta^{++}$ lineshape, a dimension in which acceptance is uncorrelated with decay angles.  We assess the systematic uncertainties in these corrections later.

To validate the statistical properties of our technique, we analyze simulated data from many toy experiments and find that our method for extracting $\Sigma$ is unbiased.  We estimate the statistical uncertainty in our measurement by examining the variance of large ensembles of toy experiments modeled to match our data.  With these uncertainties, the results from 0/90 and -45/45 data sets agree statistically with $\chi^2 /\text{NDF}$=0.35 (NDF=15).  We combine measurements from the independent 0/90 and -45/45 data sets, which have comparable statistical precision, by averaging the results.  In constructing the uncertainty on this average, we assume that individual systematic errors in the measurement technique (detailed below) are fully correlated.

To study systematic uncertainty related to choice of fitting scheme, we perform additional evaluations of $\Sigma$ while independently varying: background polynomial from fourth to eighth order, choice of fit range, whether to allow individual $\Delta^{++}$ signal parameters to float, and removal of efficiency correction to the $\Delta^{++}$ lineshape. To study the systematic uncertainty related to reliance on MC-determined corrections applied to the phase space of the $\Delta^{++}$, we perform additional evaluations of $\Sigma$ by varying the efficiency cutoff and systematically deforming the efficiency map. We also roughly describe $\Delta^*$ contributions using a double Gaussian shape, fitting to the region of $1.14$~GeV$/c^2<m(\pi^+ p)<3.50$~GeV/$c^2$ as an additional study. Each fit variation produces changes that are largely uncorrelated in $t$ and provide similar fit quality and results as the nominal scheme.  It is important to note that variations in fitting scheme often affect $Y_N$ and $Y_D$ in the same way, which reduces the dependence of the extracted value of $\Sigma$ on the fit scheme.  Nevertheless, we find that systematic uncertainties are comparable to or larger than statistical uncertainties in several regions of $t$. Other sources of uncertainty investigated include uncertainty in flux, uncertainty in polarization due to limited triplet statistics, variations in number of beam bunches selected for accidental subtraction, varying $\phi_\text{lin}$ within experimental uncertainties, and choice of binning. These potential sources of systematic uncertainty are described in greater detail in Ref.~\cite{JZ_thesis}.  The systematic uncertainty in $P_\gamma$, the polarization as measured by the TPOL, produces a relative uncertainty of 1.5\% on the {\em magnitude} of the measured value of $\Sigma$ that is fully correlated amongst all $t$ regions.

As an additional check, the analysis was repeated with varied selections of $m(\pi^+ \pi^-)$ region to include greater amounts of $\rho$ and $\Delta^*$ backgrounds into the analysis (refer to Fig.~\ref{fig:dalitz}). The same systematic variations as described above were then also repeated. We found consistent results, even when all events with $m(\pi^+ \pi^-)<1.1$~GeV/$c^2$, {\it i.e.,} all $\rho$ backgrounds, were included.

The asymmetry $\Sigma$ of the background can similarly be evaluated by inserting background yields to Eqs.~ \ref{eq:den} and \ref{eq:num}. In the mass range $1.14$~GeV$/c^2<m(\pi^+ p)<1.60$~GeV/$c^2$, the background is found to have a negative asymmetry without clearly discernible $t$ dependence.

\section{DISCUSSION OF RESULTS}

\begin{figure}
	\includegraphics[width=\linewidth]{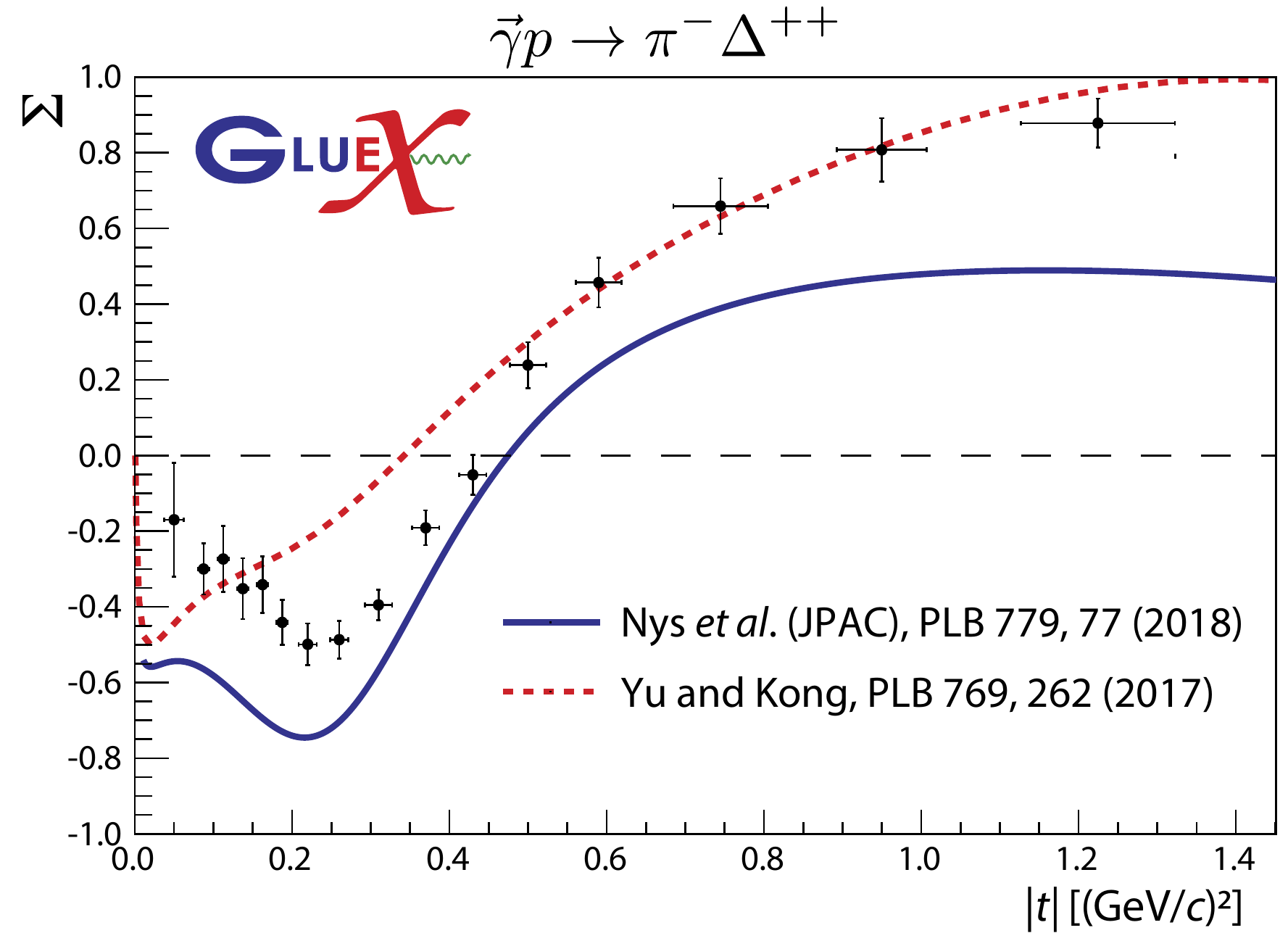}
	\caption{
		Beam asymmetry $\Sigma$ vs. $|t|$ compared to theoretical predictions. The error bars indicate the statistical and systematic uncertainties combined in quadrature.
	}
	\label{fig:piDelta_asymm}
\end{figure}

The results of beam asymmetry $\Sigma$ for $\pi^-\Delta^{++}$ photoproduction are listed in Table~\ref{tab:results} and displayed in Fig. \ref{fig:piDelta_asymm} with theoretical predictions at 8.5~GeV provided by Nys \textit{et al.}~\cite{jannes_piDelta} and B.-G. Yu and K.-J. Kong~\cite{yu_piDelta}. Several trends are apparent from the data. The asymmetry is negative in the range of approximately $|t|<0.45$~(GeV/$c$)$^2$, demonstrating that negative naturality pion exchange is favored at smaller $|t|$. In the range $|t|<0.25$~(GeV/$c$)$^2$, the asymmetry is negative and downward sloped as magnitude $|t|$ increases. This is consistent with mixed-naturality modifications to one-pion exchange, which are sharply peaked in the forward direction. For $|t|>0.45$~(GeV/$c$)$^2$ the asymmetry becomes positive, consistent with descriptions including positive naturality vector $\rho$ and tensor $a_2$ exchanges.

We find that the model of Nys \textit{et al.} describes the general shape of the asymmetry over $|t|$, though it predicts an overall lower value of $\Sigma$. The model by Yu and Kong appears to slightly better describe the asymmetry for $|t|$ larger than 0.5~(GeV/$c$)$^2$; however, it predicts a minimum value and upward rise at much lower $|t|$ than observed.

\begin{table}
\caption{
		Table of results.  The uncertainty on $|t|$ is the RMS of values in the $\Delta^{++}$ signal region.  The uncertainties on $\Sigma$ are statistical and systematic (uncorrelated across $t$ bins) respectively.  There is an additional fully correlated systematic uncertainty of 1.5\% on the magnitude of $\Sigma$.
		\label{tab:results}}
\begin{tabular}{cc}
\hline \hline
~~~$|t|$ (GeV/$c$)$^2$~~~~& $\Sigma$ \\ 
\hline
\hline
0.050 $\pm$ 0.012             & -0.17    $\pm$ 0.04         $\pm$ 0.15        \\
0.088 $\pm$ 0.007             & -0.30    $\pm$ 0.04         $\pm$ 0.06        \\
0.113 $\pm$ 0.007             & -0.27    $\pm$ 0.04         $\pm$ 0.08        \\
0.138 $\pm$  0.007            & -0.35    $\pm$ 0.04         $\pm$ 0.07        \\
0.163 $\pm$  0.007            & -0.34    $\pm$ 0.04         $\pm$ 0.06        \\
0.188 $\pm$ 0.007             & -0.44    $\pm$ 0.04         $\pm$ 0.04        \\
0.220 $\pm$ 0.011             & -0.50    $\pm$ 0.04         $\pm$ 0.04        \\
0.260 $\pm$ 0.011             & -0.49    $\pm$ 0.04         $\pm$ 0.03        \\
0.310 $\pm$ 0.017             & -0.39    $\pm$ 0.03         $\pm$ 0.02        \\
0.370 $\pm$ 0.017             & -0.19    $\pm$ 0.04         $\pm$ 0.03        \\
0.430 $\pm$ 0.017             & -0.05    $\pm$ 0.04         $\pm$ 0.04        \\
0.500 $\pm$ 0.023             & ~0.24    $\pm$ 0.03         $\pm$ 0.05        \\
0.590 $\pm$ 0.029             & ~0.46    $\pm$ 0.03         $\pm$ 0.06        \\
0.745 $\pm$ 0.060             & ~0.66    $\pm$ 0.02         $\pm$ 0.07        \\
0.950 $\pm$ 0.057             & ~0.81    $\pm$ 0.03         $\pm$ 0.08        \\
1.225  $\pm$ 0.098            & ~0.88    $\pm$ 0.04         $\pm$ 0.05        \\ 
\hline
\hline
\end{tabular}
\end{table}

In summary, we have measured the beam asymmetry $\Sigma$ as a function of $t$ for the reaction $\vec{\gamma} p \rightarrow \pi^- \Delta^{++}$ at $E_\gamma=8.5$~GeV using data from the \textsc{GlueX} experiment. These measurements are the first in this energy range and are of higher precision than and complementary to those made at higher photon beam energies~\cite{SLAC16GeV_asymm}. In the $t$-channel particle exchange picture, our measurements indicate that the naturality of exchanged Reggeons changes significantly as a function of $|t|$, consistent with pion exchange at smaller $|t|$ and natural exchange processes at higher $|t|$. These results constrain models for $t$-channel photoproduction of pions, which will be useful for understanding backgrounds in both hybrid meson searches and baryon spectroscopy studies at lower energies.

\section{ACKNOWLEDGMENTS}

This work was supported by the US Department of Energy Office of Nuclear Physics under award DE-FG02-05ER41374. We recognize additional support by the Natural Sciences and Engineering Research Council of Canada grant SAPPJ-2018-00021. We acknowledge in particular input from from V.~Mathieu to this work, as well as productive discussions with A.~Szczepaniak, J.~Nys, V.~Mathieu, M.~Mikhasenko, and B.-G.~Yu.   We would also like to acknowledge the outstanding efforts of the staff of the Accelerator and the Physics Divisions at Jefferson Lab that made the experiment possible.

This work was also supported in part by the U.S. Department of Energy, the U.S. National Science Foundation, the German Research Foundation, GSI Helmholtzzentrum f\"{u}r Schwerionenforschung GmbH, the Russian Foundation for Basic Research, the UK Science and Technology Facilities Council, the Chilean Comisi\'{o}n Nacional de Investigaci\'{o}n Cient\'{i}fica y Tecnol\'{o}gica, the National Natural Science Foundation of China, and the China Scholarship Council. This material is based upon work supported by the U.S. Department of Energy, Office of Science, Office of Nuclear Physics under contract DE-AC05-06OR23177. 

\bibliography{GlueX_pimDeltapp}

\end{document}